# ЭФФЕКТИВНЫЕ ПАРАМЕТРЫ ПОРИСТОГО СЛОЯ КРЕМНИЕВЫХ СОЛНЕЧНЫХ ЭЛЕМЕНТОВ

**З. Ж. Жанабаев, К. К. Диханбаев**

*Казахский национальный университет им. аль-Фараби,*
*г. Алматы, Казахстан*

Теоретически и экспериментально показано существование оптимальной толщины пористого слоя солнечного элемента, имеющего антиотражающее назначение. Теория учитывает совместные механизмы генерации и рекомбинации электронов и дырок, образуемых воздействием фотонов.

Пористая поверхность кремния получена электрохимическим травлением. Сканирующая электронная микроскопия, измерения тока короткого замыкания, напряжения холостого хода, времени травления показали, что при толщине пористого слоя около 2/3 диффузионной длины электрона эффективность солнечного элемента возрастает на 30 % по сравнению с ее значением при отсутствии текстурирования.

Преобразование солнечной энергии в электрическую является важной научно-технической проблемой, связанной с перспективным направлением энергетики будущего. Начиная с середины прошлого века методы разделения электронов и дырок, возникающих под действием фотонов в легированном полупроводнике, в гетероструктурах различных химических соединений интенсивно изучаются и в настоящее время [1-5].

Среди многообразия повышения эффективности солнечных элементов (СЭ) мы выделим текстурирование их поверхности с целью уменьшения отражения света [6-13]. В исследованиях такого типа электрохимическим травлением получают СЭ в виде пленок с пористой поверхностью, например, с пористым кремнием (por-Si). Размеры пор составляют порядка $10-10^3$ нм. Изменения морфологии пленок текстурированием дает антиотражающий эффект порядка 10 %, такого же порядка наблюдается повышение коэффициента полезного действия СЭ. Очевидно, что эти показатели среднестатистические. Результирующая фотоэлектродвижущая сила пленок зависит от взаимовлияния многих физических и геометрических факторов: механизмы генерации и рекомбинации электронов и дырок, длина волны эффективной части спектра солнечного излучения, топологические и метрические характеристики пористого слоя. Естественно ожидать, что возможно существование оптимальных наборов параметров, характеризующих указанные факторы. Нам неизвестны целенаправленные исследования такого плана, хотя известно множество экспериментальных фактов о роли отдельных параметров. Универсальные рекомендации по выбору оптимального набора параметров могут быть получены в результате сочетания теории с экспериментом.

Целью настоящей работы является теоретическая оценка толщины пористого слоя кремния над плоскостью $p-n$-перехода, обеспечивающей максимум генерации электронов под действием падающих фотонов с учетом их рекомбинации с дырками и сопоставление результата с экспериментом.

В легированном слое полупроводника $n$-типа имеются наноразмерные поры, которые уменьшают отражение фотонов, подающих с энергией $\hbar\omega$, длиной волны $\lambda$.

Диаметр $(d)$, высота пор $(x)$, отсчитываемая от лицевой поверхности, и расстояния между ними $(l)$ одного порядка с $\lambda$, поэтому реализуется эффект антиотражения. Толщина пористого слоя $L$ должна быть подобрана примерно равной диффузионной длине носителей зарядов $L_{Д}$ т.е. расстоянию, на котором их концентрация заметно (в $e$ раз) уменьшается. Если $L \ll L_{Д}$, то мало число выбитых фотонами избыточных электронов, если $L \gg L_{Д}$, то много рекомбинированных электронов с дырками. Задача заключается в поиске относительноного расстояния $x/L_{Д}$, на котором эти два эффекта сбалансированы. Диффузионная длина $L_{Д}$ является физическим параметром, обычно его оценивают в виде

$$L_{Д} = \sqrt{Dt}, \qquad (1)$$

где $D$ - коэффициент диффузии носителя заряда в конкретном полупроводнике, $t$ - его время жизни. Более удобной является стандартная оценка $L_{Д}$ в виде радиуса Дебая:

$$L_{Д} = V_{T}/\omega_0 = \sqrt{\frac{kT\varepsilon}{Ne^2}} = 4{,}9\sqrt{\frac{\varepsilon T(^{0}\mathrm{K})}{N(\mathrm{см}^{-3})}} см, \qquad (2)$$

где $V_{T}$ - тепловая скорость носителей заряда, $\omega_0$ - частота собственных колебаний электрона, $N$ - концентрация носителя избыточного заряда, $e$ - заряд электрона, $\varepsilon$ - относительная диэлектрическая проницаемость среды, $T(^{0}\mathrm{K})$ - абсолютная температура.

Значение $\varepsilon = 1$ для воздуха внутри пор, для кремния $\varepsilon = 11{,}8$. Число избыточных электронов и дырок генерируемых ежесекундно в единице объема полупроводника фотоном с энергией $\hbar\omega$, определяется как

$$N = \beta W/\hbar\omega, \qquad \hbar\omega \geq E_g, \qquad (3)$$

где $W$ - мощность солнечного излучения в объемном слое достаточной толщиной $1/4$ длины волны фотона $\lambda = \omega c/2\pi\sqrt{\varepsilon}$ для максимального поглощения излучения, $E_g$ – ширина энергетически запрещенной зоны полупроводника, $\beta$ - квантовый выход.

Для местоположения и суточного времени эксплуатаций СЭ измеряется интенсивность солнечного излучения $I(Bm/м^2)$. Значение $W(Bm/см^3)$ выражается через $I$:

$$W(\frac{Bm}{см^3}) = 10^6 I(\frac{Bm}{см^3})\frac{\lambda(м)}{4} \qquad (4)$$

Для обычных условий для фотовольтаики: $T = 3 \cdot 10^2 K$, $\varepsilon = 11{,}8$, $\beta = 1$, $\frac{\lambda}{4} = 10^{-7} м$, $I = 0{,}7\frac{кBm}{м^2}$, $\hbar\omega \geq E_g = 2 \cdot 10^{-19} Дж, W = 0{,}7 \cdot 10^2 \frac{Bm}{см^3}$.

Из формул (2), (3) следует оценка $L_Д \approx 10^3 нм$. Диффузионная длина составляет величину порядка микрона, является измеримым физическим параметром и дальнейшая задача стоит в определении $x_*/L_Д$, где $x_*$ – оптимальная толщина пористого слоя.

Формула (3) определяет число электронов $N$ в поверхностном слое толщиной $\lambda/4$.

Рассмотрим его изменение $N(х)$ в направлении $p-n$-перехода. Число сгенерированных электронов под действием фотонов пропорционально высоте пор $x$:
$$N_g = gpx\hbar\omega, \qquad 0 \leq p \leq 1, \qquad (5)$$
где $g$ – коэффициент генерации электронов по $x$, $p$ – коэффициент пористости пленки.

Число рекомбинированных пар электрон-дырка пропорционально как росту, так и убыванию величин $p$, $x$:
$$N_z = zp(1-p)x(L_Д - x), \qquad (6)$$
где $z$ – коэффициент рекомбинации.

Совместная вероятность реализации двух независимых статистических процессов генерации и рекомбинации
$$P(N_g, N_z) = \frac{gzp^2(1-p)}{N^2}\hbar\omega x^2(L_Д - x) \qquad (7)$$

Условие максимумов $P(N_g, N_z)$ по $p$, $x$:
$$\frac{\partial P(N_g, N_z)}{\partial x}\bigg|_{x=x_*} = 2x_*(L_Д - x_*) - x_*^2 = 0, \quad x_* = \frac{2}{3}L_Д \qquad (8)$$
$$\frac{\partial P(N_g, N_z)}{\partial p}\bigg|_{p=p_*} = 2p_*(1-p_*) - p_*^2 = 0, \quad p_* = \frac{2}{3} \qquad (9)$$

Оптимальная толщина пористого слоя составляет $2/3$ часть диффузионной длины электронов при коэффициенте пористости $p = p_* = 2/3$.

Мы ниже рассмотрим возможность наблюдения в эксперименте закономерности, описываемой формулой (8). Измерение пористости наноструктурированных пленок и проверка формулы (9) представляет отдельную экспериментальную задачу.

В качестве исходного кремния был использован монокристаллический кремний $p$-типа проводимости, толщиной 350 мкм, с удельным сопротивлением 12 Ом·см, $p-n$-переход сформирован с помощью термодиффузии фосфора. Толщина $n^+$-слоя порядка диффузионной длины электрона от 700-1500 нм, это позволяет выбрать высоту пор, образуемых после травления поверхности, меньше, чем расстояние до $p-n$-перехода. На рис. 1 показана конструкция солнечного элемента с пористым кремнием.

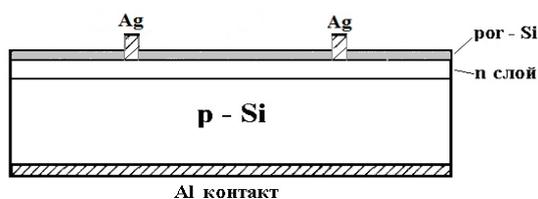

**Рис. 1.** Конструкция солнечного элемента с пористым кремнием

Контакт серебро-алюминий был получен путем напыления металлического $Ag$. Пористый кремний получен с помощью электрохимического анодирования поверхности $n^+$-слоя при различных плотности тока и времени травления. Электрохимическое анодирование проводилось в электролите $HF$: этанол в соотношении 1:1 и при плотности тока анодизации 20 $мА/см^2$.

С увеличением плотности тока анодизации и времени травления глубина проникновения пористого покрытия увеличивается в направлении к плоскости $p-n$-перехода. Для поиска эффективной толщины пористого слоя кремния $x_*$, при котором ток короткого замыкания солнечного элемента имеет максимальное значение, было проведено травление диффузионного n-слоя СЭ различной длительностью по времени.

Были получены изображения микроструктуры поперечного сечения пористого кремния, охватывающего $p-n$-переход с помощью сканирующего электронного микроскопа (СЭМ). С целью лучшей визуализации слоев $p-n$-перехода и пористой структуры использовались раствор высокой концентрации $HF$ и сильное освещение вольфрамовой лампой торца образца.

На рис. 2 а) показано СЭМ-изображение поперечного сечения образца с пористым кремнием при коротком времени травления $n$-слоя, на рис. 2 б) - при длительном травлении.

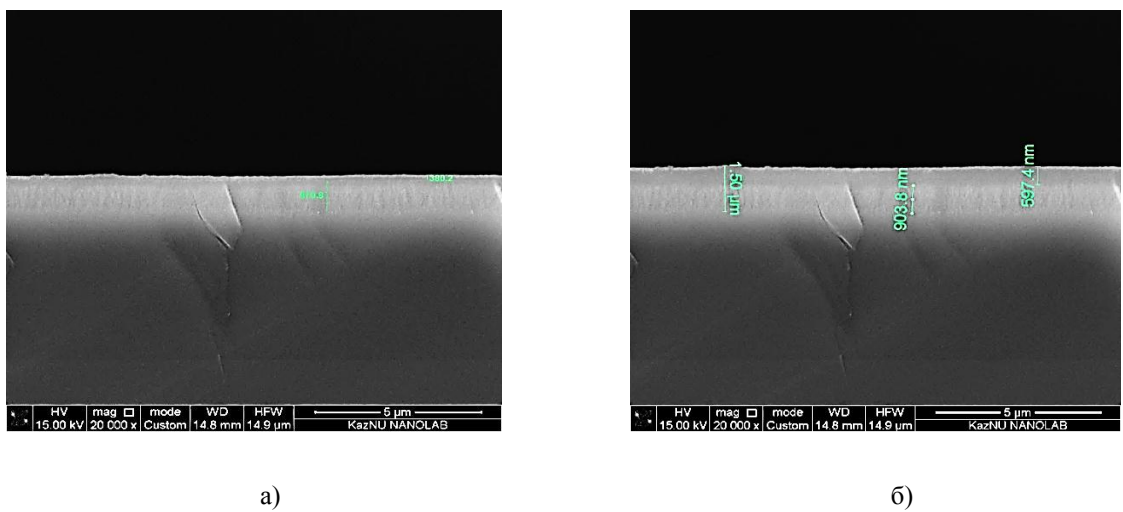

а)  б)

**Рис. 2.** а), б) - СЭМ изображения поперечного сечения СЭ с пористым кремнием

Из СЭМ-изображений видны области $p-n$-перехода, толщины $n$-слоя (1,5 мкм) и пористого слоя (903,8 нм) СЭ.

Были измерены токи короткого замыкания $I_{кз}$ образца СЭ с пористым слоем, при освещении вольфрамовой лампой накаливания мощностью 87 $мВт/см^2$. Также измерены напряжения холостого хода и коэффициент заполнения по кривой вольтамперной характеристики солнечного элемента.

Вначале были измерены выходные параметры полированного СЭ без пористого слоя, затем с различной толщины пористого кремния.

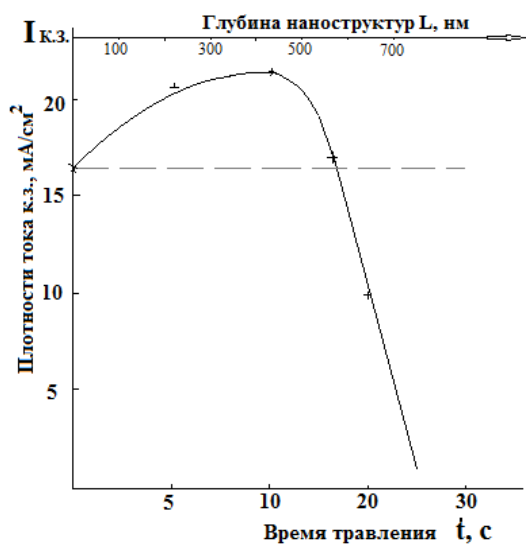

**Рис. 3.** - Изменение плотности тока короткого замыкания СЭ с ростом глубины пор (вернее шкала по оси абсцисс), пропорциональной времени травления

Рис. 3 показывает существование эффективной глубины пор, при которой плотность тока короткого замыкания СЭ имеет максимальное значение. Глубина нанопор определялась из СЭМ изображении (рис. 2 а), б)).

Максимум тока короткого замыкания 22,1 $мА/см^2$ соответствует времени травления 10 с и 420 нм глубины залегания пористого кремния в n-слое СЭ. Ток короткого замыкания увеличивается на 30 % от первоначального значения без пористого кремния.

На рис. 4 показано изменение эффективности СЭ с ростом толщины пористого $n$-слоя по направлению к $p-n$-переходу. Эффективность СЭ при глубине залегания пористого слоя 420 нм от поверхности составляет около 35 % по сравнению с первоначальным значением эффективности без пористого слоя.

Из рис. 3, 4 видно, что оптимальная высота пористого слоя СЭ, обеспечивающая максимум тока короткого замыкания ($x_{*,I}$) и коэффициента полезного действия СЭ ($x_{*,W}$) равна, соответственно $x_{*,\text{I}} = 0{,}55$, $x_{*,\text{W}} = 0{,}70$, что ближе к теоретическому значению, $x_{*,\text{W}} \approx \frac{2}{3} L_\text{Д}$ (если принять $L_\text{Д} \approx 600$ нм, согласно рис. 4).

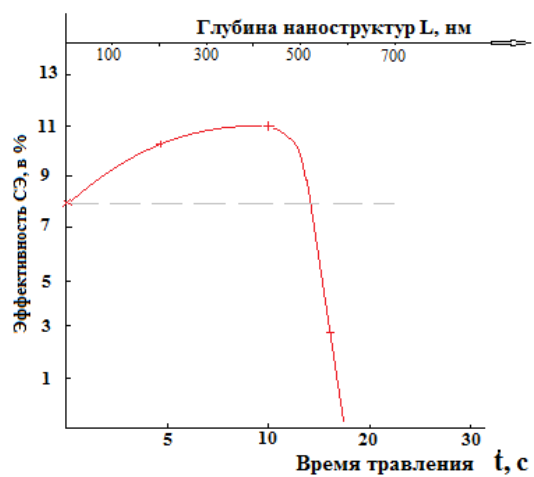

**Рис. 4.** - Изменение эффективности СЭ с ростом глубины пор в $n$-слое в направлении к плоскости $p-n$-перехода (верхняя шкала по оси абсцисс)

Широко используемая в фотовальтаике технология текстурирования поверхности солнечных элементов с целью обеспечения антиотражения, пассивации рекомбинации носителей заряда нуждается в теоретическом обосновании. В работе теоретически и экспериментально показано существование оптимальной глубины текстурирования в виде пористого слоя кремния. Из теории также следует существование оптимального значения пористости рабочей части солнечного элемента.

Результаты настоящей работы могут быть использованы для существенного (до 35 %) повышения эффективности солнечных элементов. Особенно важно то, что доказанный в настоящей работе факт существования оптимального набора параметров солнечных элементов может иметь универсальное приложение к совершенствованию различных технологий (электрохимическая порошковая технология, технологии лазерной обработки поверхности, технология использования органических элементов и т.д).